\documentclass[aps,preprintnumbers,floats, prd, twocolumn, longbibliography, nofootinbib]{revtex4-1}
\usepackage{graphicx}% Include figure files
\usepackage{dcolumn}% Align table columns on decimal point
\usepackage{bm}% bold math
\usepackage{amsmath}
\usepackage{amsthm}
\usepackage{multirow}
\usepackage{enumerate}
\usepackage{amsfonts}
\usepackage{ifthen}
\usepackage{psfrag}
\usepackage{slashed}
\usepackage{hyperref}
\usepackage{gensymb}
\usepackage[utf8]{inputenc}
\usepackage{color}
\usepackage{subfigure}
\usepackage{ulem}
\usepackage[utf8]{inputenc}
\usepackage{float}

\newcommand{\be}{\begin{equation}}
\newcommand{\ee}{\end{equation}}
\newcommand{\bea}{\begin{eqnarray}}
\newcommand{\eea}{\end{eqnarray}}

\begin{document}

\title{Probing Dark matter-Baryon Interaction with S301\\---the Fastest Known Star in the Milky Way}

\author{Yugen Lin$^{1}$}
\email{linyugen@itp.ac.cn}

\affiliation{$^1$ Institute of Theoretical Physics, Chinese Academy of Sciences, Beijing, 100190, China}

\begin{abstract}
The recent discovery of S301 provides an unique opportunity to investigate dark matter-baryon interaction in the innermost region of the Milky Way. S301 follows a highly eccentric orbit around Sgr~A*, and its pericenter distance is only around 12 AU with velocity about $25000\,{\rm km/s}$. These extreme properties make S301 highly sensitive to dark matter-baryon scattering in the center of galaxy, where the dark matter density may be strongly enhanced. We have calculated the orbit-averaged energy deposited rate in S301 via dark matter-proton scattering, and show that over a broad dark matter parameter space, cross sections satisfying the existing bounds from terrestrial experiments and astronomical observations can inject sufficient energy to modify the luminosity and evolution of S301. Our results establish S301 as an exceptionally sensitive probe of dark matter-baryon interactions and open a new avenue for probing dark matter interactions in high-density and high-velocity astrophysical environments.
\end{abstract}

\maketitle
%%%%%%%%%%%%%%%%%%%%%%%%%%%%%%%%%%%%%%%%%%%%%%%%%%%%%%%%%%%%%%%%%%%%%
%%%%%%%%%%%%%%%%%%%%%%%%%%%%%%%%%%%%%%%%%%%%%%%%%%%%%%%%%%%%%%%%%%%%%

\section{Introduction}
\label{sect:intro}
A wide range of cosmological and astrophysical observations provide strong evidence for the existence of dark matter (DM), which constitutes most of the matter content of the Universe~\cite{Planck:2018vyg,Arbey:2021gdg}. However, its microscopic nature and possible interactions with Standard Model (SM) particles still remain unknown. Elastic scattering between DM and nuclei is one of the most important channels to investigate DM. Over the past few decades, terrestrial direct-detection experiments have placed stringent constraints on such interactions~\cite{LZ:2024zvo,PandaX:2024qfu,XENON:2023cxc,XENON:2018voc,LUX:2016ggv,DEAP:2019yzn,PICO:2019vsc,DarkSide:2018bpj}. On the other hand, astrophysical environments, where densities, temperatures, and target sizes can differ greatly from laboratory conditions, therefore provide an important and complementary approach for probing DM interactions with ordinary matter~\cite{Baryakhtar:2017dbj,Graham:2018efk,Leane:2020wob,Wadekar:2019mpc,Bringmann:2018cvk,Dvorkin:2013cea,IceCube:2021xzo,Lin:2025mez,Bi:2021njb,Meighen-Berger:2026idy}.

Stars in the Galactic Center around the supermassive black hole Sgr~A*, commonly known as S-stars, provide a natural laboratory to investigate physics under extreme conditions. Over the past two decades, precision astrometric and spectroscopic observations of S-stars have established the mass and geometric distance of the Sgr~A*~\cite{Schodel:2002py,Gravity:2019nxk}, and enabled detections of gravitational redshift, Schwarzschild precession, constraints on fifth forces and extended mass distributions~\cite{GRAVITY:2018ofz,GRAVITY:2020gka,Hees:2017aal,GRAVITY:2021xju}.

The star S301, recently discovered by the GRAVITY+ Collaboration~\cite{Dayem:2026ktt}, may offer an unique opportunity to investigate the DM-baryon interaction.  S301 is a faint main-sequence star with a mass approximately $1.5\,M_\odot$ and a radius of about $1.4\,R_\odot$, and follows a highly eccentric orbit around Sgr~A*. In particular, it only has a pericenter distance of approximately 12 AU from Sgr~A* and a peak orbital velocity of around $25000$~${\rm km/s}$, corresponding to more than eight percent of the speed of light. S301 consequently has the smallest pericenter distance from Sgr~A* among currently known stellar, and its velocity also makes it the fastest known star in the Milky Way.

\begin{figure}[!htbp]
    \centering \includegraphics[width=\columnwidth]{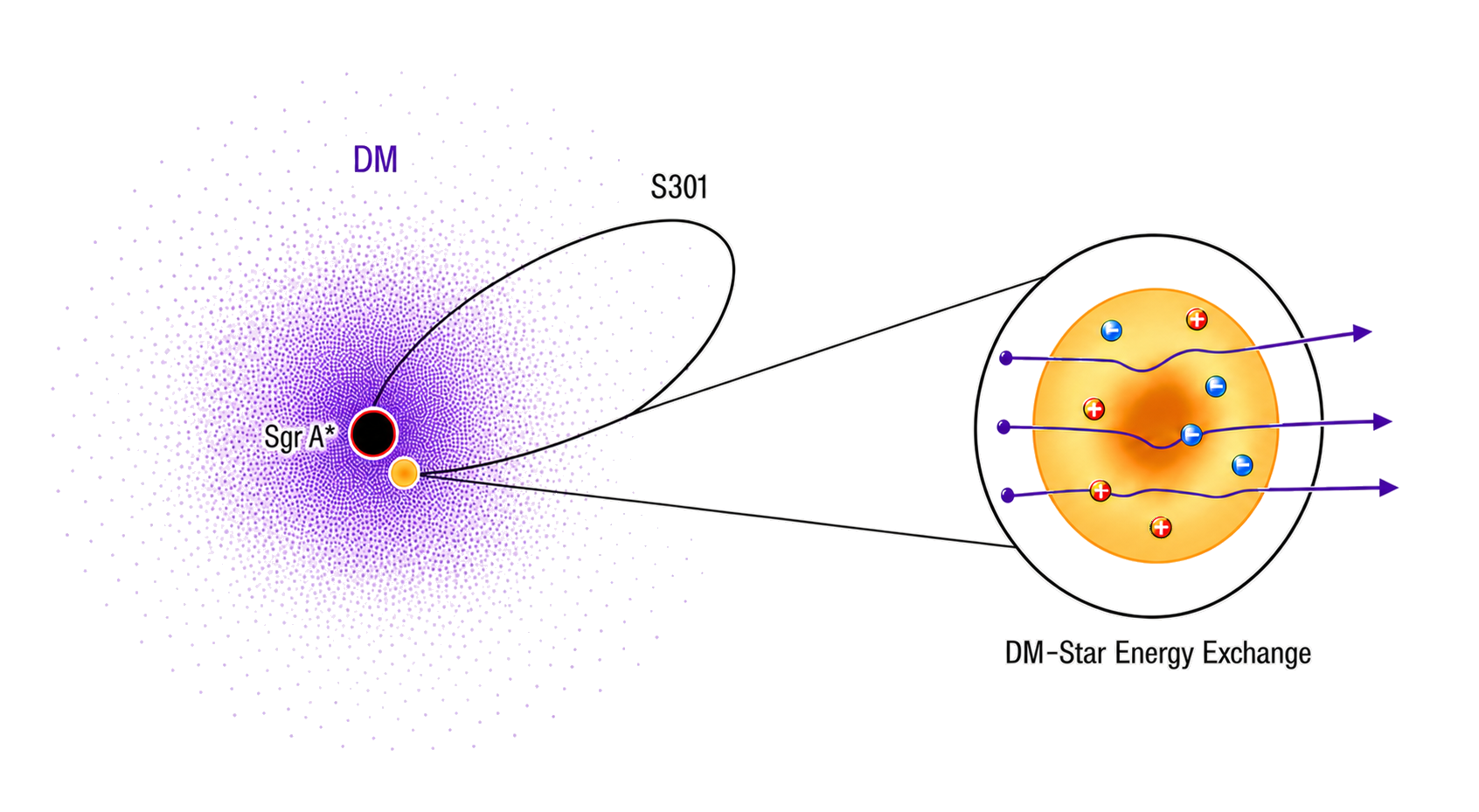}
    \caption{This schematic illustrates a star orbiting Sgr A* on a highly eccentric orbit passing through an overdense dark matter region, where the elastic scattering of dark matter with stellar can yield a net energy transfer.}
    \label{fig:illustration}
\end{figure}

In the stellar rest frame, the surrounding dark matter particles also collide with the star at high velocity and exchange energy with stellar matter. The Schematic illustration of such process is shown in Fig.~\ref{fig:illustration}. The efficiency of energy exchange depends on the DM density and the relative velocity between the star and the surrounding DM population. Both quantities can become exceptionally large near a supermassive black hole. In particular, the adiabatic growth of a black hole inside a pre-existing DM cusp may compress the surrounding halo and generate a steep density enhancement, commonly referred to as a DM spike~\cite{Quinlan:1994ed,Sadeghian:2013laa}. The density in such a spike rises rapidly toward the black hole, potentially reaching values many orders of magnitude larger than the local DM density.

When a DM spike is present, S301 passes through the highest DM density region sampled by any known star. Its exceptionally large orbital velocity further increases both the rate of DM particles crossing the star and the energy exchange rate in DM-baryon collisions. The combination of high DM density, large relative velocity, and a comparatively low-luminosity main-sequence star makes S301 a particularly sensitive target for studying DM-induced stellar heating.

In this work, we have investigated the energy exchanged between DM and the baryonic constituents of S301 through elastic scattering. We use the newest measured orbital parameters of S301 to evaluate the energy transfer rate and obtain the corresponding orbit-averaged contribution to the stellar energy budget. We compare the DM-induced energy exchange rate with the intrinsic stellar luminosity, combining the relevant parameter space with existing laboratory and astrophysical constraints. Our analysis is intended to determine which combinations of DM mass, scattering cross section can produce a non-negligible modification of the energy budget of S301. In this way, we show that S301 can provide a new probe of DM-baryon interactions in the innermost observable region surrounding the Milky Way's central supermassive black hole.

This paper is organized as follows. In
Section~\ref{sect:scattering}, we introduce the galactic DM density profiles, and present the analytic formalism for DM-induced energy exchange rate. We show our results in Section~\ref{sect:Result} and summarize our conclusions in Section~\ref{sect:conclusion}.

\section{Heating Rate via DM-baryon Scattering}
\label{sect:scattering}

In this section, we will calculate the energy transfer rate induced by DM-baryon scattering in S301. The heating rate is mainly controlled by two physical quantities: the DM density and the relative velocity between the star and DM particle. Both quantities increase rapidly toward Sgr~A$^*$, so that the energy-transfer rate is strongly enhanced, especially close to the pericenter.

For the DM distribution, the generalized NFW profile~\cite{Navarro:1995iw} is written as
\begin{equation}
    \rho_{\rm NFW}(r) = \frac{\rho_s}
    {(r/r_s)^\gamma(1+r/r_s)^{3-\gamma}},
    \label{eq:nfw_profile}
\end{equation}
where $r_s$ is the scale radius, $\rho_s$ is the scale density, and
$\gamma$ specifies the inner slope. We take the fiducial value $r_s=20~{\rm kpc}$ which is the scale radius for the Milky Way~\cite{Cirelli:2010xx}, and normalize the profile to the local DM density
$\rho_\odot=0.4~{\rm GeV\,cm^{-3}}$ at
$r_\odot=8.5~{\rm kpc}$. For the slope $\gamma$, $0.5 \lesssim\gamma \lesssim 1.5$ are well motivated~\cite{Gnedin:2003rj,Iocco:2016itg,Hooper:2016ggc,Baumgart:2025dov}. In this paper, we set $\gamma=1$, and the corresponding $\rho_s \simeq 0.35~{\rm GeV\,cm^{-3}}$.

The growth of a supermassive black hole can adiabatically compress a pre-existing DM cusp and produce a much steeper central distribution, usually referred to as a DM spike. For an initial density profile with inner slope $\gamma$, the resulting spike slope is~\cite{Gondolo:1999ef}
\begin{equation}
    \gamma_{\rm sp}
    =
    \frac{9-2\gamma}{4-\gamma}.
    \label{eq:spike_slope}
\end{equation}
The spike profile is matched continuously to the outer NFW halo at the spike radius $R_{\rm sp}$,
\begin{equation}
    \rho_{\rm sp}(r)
    =
    \rho_{\rm NFW}(R_{\rm sp})
    \left(\frac{R_{\rm sp}}{r}\right)^{\gamma_{\rm sp}},
    \qquad r\leq R_{\rm sp}.
    \label{eq:spike_profile}
\end{equation}
where $R_{\rm sp}=0.3~{\rm pc}$~\cite{Quinlan:1994ed,Sadeghian:2013laa}. For the canonical NFW slope
$\gamma=1$, Eq.~\eqref{eq:spike_slope} gives
$\gamma_{\rm sp}=7/3$. The complete spiked profile is therefore
\begin{equation}
    \rho_\chi(r)
    =
    \begin{cases}
        \rho_{\rm sp}(r), & r\leq R_{\rm sp},\\[1mm]
        \rho_{\rm NFW}(r), & r>R_{\rm sp}.
    \end{cases}
    \label{eq:composite_profile}
\end{equation}
Considering the orbit of S301 lies deep inside the adopted spike radius, it therefore makes S301 an especially sensitive probe of DM-baryon scattering in the innermost region of the galaxy.

\subsection{Energy transfer from the bulk relative motion}
\label{subsec:bulk_heating}

We now derive the energy transfer rate between the DM and stellar protons. In the galactic frame, the velocity of proton in S301 can be written as
\begin{equation}
    \mathbf{v}_{\rm proton}
    =
    \mathbf{v}_\star+\mathbf{u},
    \label{eq:proton_velocity}
\end{equation}
where $\mathbf{v}_\star$ is the bulk orbital velocity of the star and $\mathbf{u}$ is the proton thermal velocity relative to the stellar. Considering the stellar core temperature is only $T\sim 10^7$ K, i.e. about 1 keV, the thermal velocity of proton is much smaller than the orbital velocity of the star. Therefore, We begin with the limit in which the thermal velocity is neglected,
$\mathbf{u}=0$, and we assume that the DM is at rest in the galactic frame, $\mathbf{v}_\chi=0$. The relative velocity is then
\begin{equation}
    \mathbf{w}
    =
    \mathbf{v}_\chi-\mathbf{v}_{\rm proton}
    =
    -\mathbf{v}_\star.
    \label{eq:stationary_dm_relative_velocity}
\end{equation}
Equivalently, in the stellar rest frame, the protons have no bulk motion and the star is crossed by a DM wind of speed $v_\star$. This provides the leading contribution because the orbital velocity of S301 near pericenter is much larger than the characteristic thermal velocity of stellar protons.

For a DM particle of mass $m_\chi$ scattering from a
proton of mass $m_p$, the collision rate per proton is
\begin{equation}
    \Gamma_p
    =
    n_\chi w\,\sigma_{\chi p},
    \label{eq:collision_rate}
\end{equation}
where $n_\chi=\rho_\chi/m_\chi$ and $\sigma_{\chi p}$ is the
DM-proton scattering cross section. The mean proton recoil energy
is
\begin{equation}
    \left\langle T_p\right\rangle
    =
    \frac{\mu_{\chi p}^2}{m_p}w^2,
    \qquad
    \mu_{\chi p}
    =
    \frac{m_\chi m_p}{m_\chi+m_p}.
    \label{eq:mean_recoil}
\end{equation}
The energy-transfer rate per proton follows directly from the product of the collision rate and the energy transferred per collision:
\begin{align}
    \dot Q_p^{(0)}
    &=
    \Gamma_p\left\langle T_p\right\rangle
    \nonumber\\
    &=
    \rho_\chi\,\sigma_{\chi p}
    \frac{m_\chi m_p}{(m_\chi+m_p)^2}w^3.
    \label{eq:leading_heating_rate}
\end{align}
The superscript $(0)$ represents the approximation in which the
proton thermal motion has been neglected.

Equation~\eqref{eq:leading_heating_rate} displays the strong velocity dependence, where the scattering rate is proportional to $w$ and the recoil energy is proportional to $w^2$. The heating rate therefore scales as $w^3$. The mass-dependent factor $\frac{m_\chi m_p}{(m_\chi+m_p)^2}$ describes the efficiency of energy transfer in a two-body collision. It is largest when $m_\chi$ and $m_p$ are comparable and decreases when either particle is much heavier than the other.

The total instantaneous luminosity deposited in the star is
\begin{align}
    L_\chi^{(0)}(r)
    &=
    \int_{\rm star}
    n_p(\mathbf{x})\dot Q_p^{(0)}(r)\,d^3x
    \nonumber\\
    &=
    N_H\rho_\chi(r)\sigma_{\chi p}
    \frac{m_\chi m_p}{(m_\chi+m_p)^2}v_\star^3(r),
    \label{eq:instantaneous_heating}
\end{align}
where $N_H = \int_{\rm star}n_p(\mathbf{x})\,d^3x$ is the total number of hydrogen targets. The external DM density and relative velocity are approximately constant on the scale of stars, so they can be taken outside the volume integral.

\subsection{Correction from the proton thermal velocity}
\label{subsec:thermal_correction}

We next restore the proton thermal motion while continuing to treat the DM as rest in the galactic frame. In the stellar rest frame, the DM has a fixed bulk velocity $\mathbf{V}=-\mathbf{v}_\star$, while the proton has thermal velocity $\mathbf{u}$. The relative velocity and centre-of-mass velocity are respectively
\begin{equation}
    \mathbf{w}=\mathbf{V}-\mathbf{u},
    \qquad
    \mathbf{V}_{\rm CM}
    =
    \frac{m_\chi\mathbf{V}+m_p\mathbf{u}}
    {m_\chi+m_p}
    \label{eq:thermal_kinematics}
\end{equation}
The proton energy change after collision is
\begin{equation}
    \left\langle\Delta E_p\right\rangle
    =
    \mu_{\chi p}\,
    \mathbf{V}_{\rm CM}\cdot\mathbf{w}.
    \label{eq:angular_average}
\end{equation}
The scalar product appearing here can be expressed as
\begin{equation}
    \mathbf{V}_{\rm CM}\cdot\mathbf{w}
    =
    \frac{
        m_\chi V^2
        +(m_p-m_\chi)\mathbf{V}\cdot\mathbf{u}
        -m_pu^2}
    {m_\chi+m_p}.
    \label{eq:vcm_dot_w}
\end{equation}
For an isotropic Maxwell distribution, $\left\langle\mathbf{u}\right\rangle=0$, $\left\langle u^2\right\rangle=\frac{3k_{\rm B}T_p}{m_p}$, and hence $\left\langle\mathbf{V}_{\rm CM}\cdot\mathbf{w}\right\rangle=\frac{m_\chi V^2-3k_{\rm B}T_p}{m_\chi+m_p}$.
The thermally averaged energy transferred to the proton is therefore
\begin{equation}
    \left\langle\Delta E_p\right\rangle
    =
    \frac{m_\chi m_p}{(m_\chi+m_p)^2}
    \left(m_\chi V^2-3k_{\rm B}T_p\right).
    \label{eq:thermal_energy_transfer}
\end{equation}

This result separates the two competing physical effects. The term
$m_\chi V^2$ describes heating driven by the relative bulk motion,
whereas $3k_{\rm B}T_p$ represents the thermal energy transfer of the stellar protons. When the first term dominates, energy flows from the moving DM population into the star. Combining the thermal contribution, the corresponding energy-transfer rate can then be written as
\begin{equation}
    \dot Q_p^{(T)}
    =
    \rho_\chi\sigma_{\chi p}
    \frac{m_\chi m_p}{(m_\chi+m_p)^2}
    V^3
    \left(
        1-\frac{3k_{\rm B}T_p}{m_\chi V^2}
    \right).
    \label{eq:thermal_corrected_rate}
\end{equation}
Thus, the relative size of the temperature correction is governed by $3k_{\rm B}T_p/(m_\chi V^2)$. For S301, however, its extremely large pericenter velocity makes the bulk-motion term dominant in the mass range considered in this paper. Specifically, for $m_\chi\gtrsim10~{\rm MeV}$, the thermal motion can only bring about less than $4\%$ correction.

\subsection{Correction from the dark-matter velocity}
\label{subsec:dm_velocity}

We now relax the assumption that the DM is at rest in the
galactic frame. At the pericenter of S301, we approximate its magnitude by the local circular velocity,
\begin{equation}
    v_\chi
    \simeq
    v_{\rm circ}
    =
    \sqrt{\frac{GM_{\rm BH}}{r_p}}.
    \label{eq:dm_velocity}
\end{equation}
The stellar pericenter velocity follows from the Keplerian orbit,
\begin{equation}
    v_{\rm peri}^2
    =
    (1+e)v_{\rm circ}^2.
    \label{eq:stellar_pericenter_velocity}
\end{equation}
We use $\theta$ to denote the angle between the stellar and DM velocity, their relative velocity is
\begin{equation}
    w
    =
    \left|
        \mathbf{v}_{\rm peri}-\mathbf{v}_\chi
    \right|
    =
    \sqrt{
        v_{\rm peri}^2+v_\chi^2
        -2v_{\rm peri}v_\chi\cos\theta
    }.
    \label{eq:relative_velocity}
\end{equation}
Because the energy-transfer rate scales as $w^3$, after averaging over the velocity direction, we can obtain
\begin{align}
    \left\langle w^3\right\rangle_\theta
    &=
    \frac{1}{2}
    \int_{-1}^{1}
    \left(
        v_{\rm peri}^2+v_\chi^2
        -2v_{\rm peri}v_\chi\mu
    \right)^{3/2}d\mu
    \nonumber\\
    &=
    \frac{
        (v_{\rm peri}+v_\chi)^5
        -|v_{\rm peri}-v_\chi|^5}
        {10v_{\rm peri}v_\chi}.
    \label{eq:w3_average}
\end{align}
Substituting Eqs.~\eqref{eq:dm_velocity} and
\eqref{eq:stellar_pericenter_velocity}, we define the velocity
correction factor
\begin{equation}
    F
    \equiv
    \frac{\left\langle w^3\right\rangle_\theta}
    {v_{\rm peri}^3}
    =
    1+\frac{2}{1+e}
    +\frac{1}{5(1+e)^2}.
    \label{eq:velocity_factor}
\end{equation}
Accordingly, the relative-velocity entering the pericenter
heating rate is
\begin{equation}
    w_p^3
    \equiv
    \left\langle w^3\right\rangle_\theta
    =
    Fv_{\rm peri}^3.
    \label{eq:wp_definition}
\end{equation}
For the highly eccentric orbit of S301, this correction is small and approximately doubles the heating rate relative to the DM at rest.

\subsection{pericenter factor and orbit-averaged heating}
\label{subsec:pericenter_factor}

The preceding expressions describe the instantaneous heating rate at a given orbital position. Considering that the DM density and star velocity reach their largest values at pericenter, we therefore introduce a pericenter factor $f_{\rm peri}$,
\begin{equation}
    \overline{L}_\chi
    =
    \frac{L_{\rm peri}\Delta t_{\rm peri}}
    {T_{\rm orb}}
    =
    f_{\rm peri}L_{\rm peri}.
    \label{eq:pericenter_factor_definition}
\end{equation}
Here $L_{\rm peri}$ is the instantaneous heating rate evaluated at
$r_p$, and $\Delta t_{\rm peri}$ is the time spent in the region where the heating is concentrated.

To specify this region, we introduce a dimensionless parameter
$f_r$ and define
\begin{equation}
    r\leq r_c,
    \qquad
    r_c=(1+f_r)r_p.
    \label{eq:pericenter_region}
\end{equation}
The choice of $f_r$ can be motivated by the radial dependence of the energy deposition. Inside the spike,
$\rho_\chi\propto r^{-\gamma_{\rm sp}}$, while the gravitational
potential gives $v\propto r^{-1/2}$. The instantaneous heating
therefore scales as
\begin{equation}
    L_\chi(r)
    \propto
    \rho_\chi(r)v^3(r)
    \propto
    r^{-\gamma_{\rm sp}-3/2}.
    \label{eq:radial_luminosity_scaling}
\end{equation}
Using $dQ=L_\chi dt$ and $dt\simeq dr/v$, the accumulated energy satisfies
\begin{equation}
    \frac{dQ}{dr}
    \propto
    \frac{L_\chi}{v}
    \propto
    r^{-(\gamma_{\rm sp}+1)}.
    \label{eq:radial_energy_scaling}
\end{equation}
The fraction of the heating deposited between $r_p$ and
$r_c=(1+f_r)r_p$ is consequently estimated as
\begin{equation}
    \frac{Q(r\leq r_c)}{Q_{\rm total}}
    \simeq
    1-(1+f_r)^{-\gamma_{\rm sp}}.
    \label{eq:heating_fraction}
\end{equation}
For $\gamma_{\rm sp}=7/3$ and $f_r=2$, this fraction is about $0.923$. We therefore take $r_c=3r_p$, which captures approximately $92\%$ heating contribution in star orbit.

The corresponding time fraction in the near-pericenter region can be calculated  from the Keplerian orbit. In terms of the eccentric anomaly $E$, the radial distance is $r(E)=a(1-e\cos E)$. The eccentric anomaly at the boundary of the pericenter region is determined from
\begin{equation}
    a(1-e\cos E_{\max})
    =
    r_p(1+f_r),
\end{equation}
which gives
\begin{equation}
    E_{\max}
    =
    \arccos\left[
        1-\frac{f_r(1-e)}{e}
    \right].
    \label{eq:Emax}
\end{equation}
Because the near-pericenter region is symmetric about pericenter, the total time spent inside it is twice the time between $E=0$ and
$E=E_{\max}$. The orbital time fraction is therefore
\begin{equation}
    f_{\rm peri}\equiv
    \frac{\Delta t_{\rm peri}}{T_{\rm orb}}
    =
    \frac{E_{\max}-e\sin E_{\max}}{\pi}.
    \label{eq:fperi}
\end{equation}

The orbit-averaged DM-induced luminosity can now be written as
\begin{equation}
    \overline{L}_\chi
    =
    f_{\rm peri}N_H
    \rho_\chi(r_p)
    \sigma_{\chi p}
    \frac{m_\chi m_p}{(m_\chi+m_p)^2}
    w_p^3,
    \label{eq:final_heating}
\end{equation}
where $\rho_\chi(r_p)$ is the DM density at pericenter, $w_p^3$ provides the enhancement effect at high velocity, and $f_{\rm peri}$ represents the duration of heating at the near-pericenter region.

\section{Result and Discussion}
\label{sect:Result}

In this section, we will calculate the energy transfer rate using the S301 observation data~\cite{Dayem:2026ktt}. The stellar and orbital parameters are summarized in Table~\ref{tab:S301_parameters}. Due to the lack of radial velocity information, the astrometric data admit two
possible three-dimensional orientations of the orbit. The two solutions differ substantially in
the orientation angles $(i,\Omega,\omega)$, but have nearly identical
semi-major axes, eccentricities, orbital periods, pericenter distances,
and pericenter velocities. The heating rate is therefore only weakly affected by this degeneracy.

\begin{table*}[t]
\centering
\renewcommand{\arraystretch}{1.2}
\resizebox{\textwidth}{!}{%
\begin{tabular}{lcccccccc}
\hline\hline
Orbit
& $a\,[{\rm AU}]$
& $e$
& $i\,[^\circ]$
& $\omega\,[^\circ]$
& $\Omega\,[^\circ]$
& $T_{\rm orb}\,[{\rm yr}]$
& $r_p\,[{\rm AU}]$
& $v_{\rm peri}\,[{\rm km\,s^{-1}}]$ \\
\hline
S301-I
& $687.0\pm5.8$
& $0.9832\pm0.0010$
& $124.09\pm1.10$
& $293.4\pm2.2$
& $73.8\pm3.5$
& $8.68\pm0.11$
& $11.54\pm0.69$
& $25\,600\pm800$ \\
S301-II
& $687.0\pm5.8$
& $0.9824\pm0.0011$
& $122.84\pm1.12$
& $115.1\pm2.0$
& $256.9\pm3.3$
& $8.68\pm0.11$
& $12.09\pm0.76$
& $25\,000\pm800$ \\
\hline\hline
\end{tabular}%
}
\caption{Stellar and orbital parameters for S301~\cite{Dayem:2026ktt}. Here, $a$ is the orbital semimajor axis, $e$ is the eccentricity, $i$ is the orbital inclination, $\omega$ is the orientation of the orbital ellipse, $\Omega$ is the longitude of the ascending node, $T_{\rm orb}$ is the orbital period, $r_p$ is the pericenter distance, and $v_{\rm peri}$ is the stellar velocity at pericenter. The two rows correspond to the two allowed orbital orientations.
Solution I is used as the fiducial configuration in our analysis. The stellar mass, radius, and luminosity are respectively $M_\star=1.5\,M_\odot$, $R_\star=1.4\,R_\odot$, and $L_\star=5.5\,L_\odot$. }
\label{tab:S301_parameters}
\end{table*}

For the stellar composition, we take a hydrogen mass fraction $X_H=0.7$, giving $ N_H=\frac{X_HM_\star}{m_p} \simeq1.25\times10^{57}$. For the fiducial $\gamma=1$ halo, the DM density at the pericenter of Solution~I is $\rho_{\rm sp}(r_p)\simeq1.16\times10^{13}~{\rm GeV\,cm^{-3}}$. The second orbital solution gives an energy transfer rate approximately $10\%$ smaller than the first. This difference is small and we consequently use the Solution~I in the remainder of the analysis.

\begin{figure}[!htbp]
    \centering \includegraphics[width=\columnwidth]{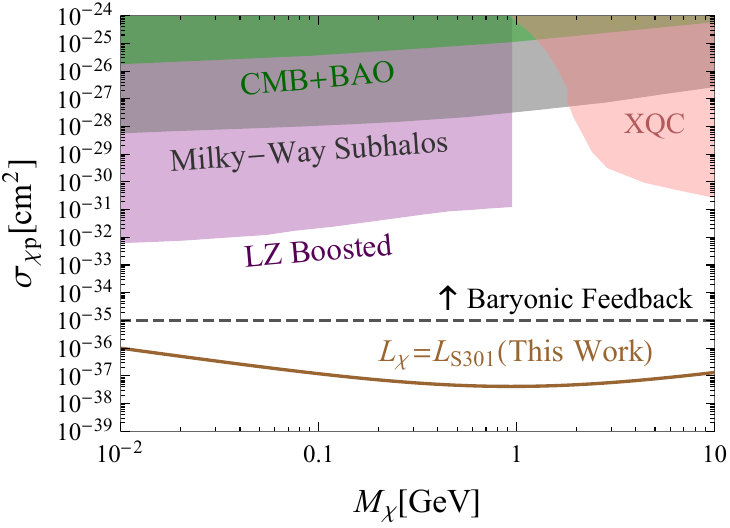}
    \caption{Constraints on elastic dark matter-proton scattering in the ($M_\chi$, $\sigma_{\chi p}$) plane. The solid brown line denotes the cross section for which the orbit-averaged dark-matter heating rate equals the luminosity of S301, using the S301-I orbit parameters. Cross sections above this curve would inject sufficient energy to produce an ${\cal O}(1)$ modification of the stellar energy budget. The green, gray, purple, and rink shaded regions show existing exclusions from CMB+BAO observations, Milky-Way subhalo abundances, the LZ boosted dark matter search, and XQC experiments, respectively. The dashed horizontal line marks the approximate threshold above which baryonic feedback may modify the dark matter spike.}
    \label{fig:result}
\end{figure}

To translate the heating rate induced by dark matter collisions into a constraint on the scattering cross section, we require that the orbit-averaged energy-deposition rate satisfy $\overline{L}_{\chi}\lesssim L_{\star}=5.5\,L_\odot$. This criterion follows directly from the stellar energy budget: if the energy continuously supplied by dark matter-baryon scattering were comparable to or larger than the luminosity of the star, the additional heating could no longer be treated as a small perturbation. Over many orbital periods, it would modify the thermal equilibrium of S301 and could produce appreciable changes in its radius, effective temperature, luminosity, and main-sequence lifetime. We therefore define $\overline{L}_{\chi}=L_{\star}$ as the characteristic sensitivity threshold, and the corresponding constraints are shown in Fig.~\ref{fig:result}. This criterion provides a conservative estimate of the constraint on the dark matter-baryon scattering cross section. A more robust bound would require incorporating the energy transfer rate into a stellar-evolution code and studying the resulting structural and evolutionary response of S301, which is beyond the scope of this work.

The equal-luminosity contour (brown solid line) probes a region complementary to existing laboratory and cosmological searches. Below a few GeV masses, conventional nuclear-recoil experiments rapidly lose sensitivity because the recoil energy falls below the detector threshold. Searches for cosmic ray boosted DM partly overcome this limitation: collisions with galactic cosmic rays accelerate a small
fraction of the DM to detectable energies. The recent LZ experiment~\cite{LZ:2025iaw}, using a $4.2$ tonne-year exposure, constrains the DM-nucleon cross section down to approximately $6\times10^{-33}~{\rm cm^2}$ in the sub-GeV region, which is shown as purple shaded region. This bound remains several orders of magnitude above the S301 equal-luminosity contour.

Cosmological and structure formation constraints provide additional, although generally weaker constraints, comparisons in the mass range of interest. DM-baryon scattering suppresses the matter power spectrum on small scales and the formation of low-mass subhaloes, so the observed Milky Way satellite population can be used to constrain the DM-baryon cross section~\cite{Kennedy:2013uta,DES:2020fxi,Nadler:2019zrb,Boehm:2004th}. On the other hand, the exchange of heat between DM and baryon can lead to recombination occurring earlier, which can directly impact the CMB~\cite{Ali-Haimoud:2015pwa,Ali-Haimoud:2021lka,Xu:2018efh,Chen:2002yh}. The constraints from the datasets of CMB+BAO and Milky-Way subhalos are shown as gray and green shaded region respectively, which
are adapted from the recent study Ref.~\cite{Buen-Abad:2021mvc}. We also include bounds from the XQC rocket experiment~\cite{McCammon:2002gb,Erickcek:2007jv,Mahdawi:2018euy}, which is shown as pink shaded region. These comparisons reveal a substantial region, in which existing experimental constraints permit a DM-proton cross section large enough to provide an ${\cal O}(1)$ contribution to the luminosity of S301.

The preceding calculation assumes that the central spike is already present and is not substantially altered by the DM-baryon interaction. This assumption also need to be checked self-consistently. Repeated scattering with stars and gas can transfer energy and momentum to tightly bound DM particles, moving them onto less bound orbits and gradually reducing the central density. A simple estimate is the mean number of DM-proton collisions over the age scale of galaxy,
\begin{equation}
    P_{\chi p}
    \simeq
    t_{\rm age}\,n_p^{\rm GC}\,v_p^{\rm GC}\,
    \sigma_{\chi p},
    \label{eq:feedback_probability}
\end{equation}
where $n_p^{\rm GC}$ and $v_p^{\rm GC}$ are effective baryonic density and velocity scales characterizing the Galactic Centre environment. Requiring that a typical DM particle experiences fewer than one collision during the the age scale of galaxy gives $t_{\rm age}\,n_p^{\rm GC}\,v_p^{\rm GC}\,\sigma_{\chi p}\lesssim1$. The baryon density in the Galactic Center is $n_p^{\rm GC}\sim10^7\text{-}10^8~{\rm cm^{-3}}$~\cite{Genzel:2010zy,Schoedel:2008ny,Baganoff:2001ju}.  Taking $t_{\rm age}\simeq10~{\rm Gyr}$, $v_p^{\rm GC}\lesssim0.1c$,  we can obtain the feedback scale $\sigma_{\chi p}\lesssim10^{-35}{\rm cm^2}$. The S301 equal-luminosity contour lies below this value throughout the DM mass region considered in this paper. For example,  $m_\chi=10~{\rm MeV}$, $\sigma_{\chi p}\simeq 10^{-36}~{\rm cm^2}$,
corresponding to $P_{\chi p}\sim0.1$. Near $m_\chi=m_p$, the equal-luminosity cross section further decline and give $P_{\chi p}\sim4\times10^{-3}$. Consequently, the cross sections relevant for the S301 sensitivity do not significantly modify the formation and survival of the DM spike, and a detailed treatment of baryonic feedback would only mildly affect our conclusions.

\section{Summary}
\label{sect:conclusion}

In this work, we have investigated the effects of dark matter-proton scattering on the luminosity of star. S301 has the smallest known pericenter distance and the highest orbital velocity among the stars orbiting Sgr~A*, which makes it a particularly sensitive probe of dark matter in the innermost region of the Milky Way. For a dark-matter spike in the Galactic Center, we find that  satisfying the existing bounds from terrestrial experiments and astronomical observations, DM-baryon scattering can produce a heating rate comparable to, or even larger than, the intrinsic luminosity of S301 over a broad dark matter parameter space. Therefore, the newly discovered S301 would provide a new and unique probe of dark matter at the Galactic Center.

The framework established in this paper can be naturally extended to interactions with non-trivial velocity dependencies. For a momentum-transfer cross section $\sigma=\sigma_0 v^n$ with positive index $n$, the deposited power scales as $\overline{L}_{\chi}\propto\rho_\chi\sigma_0v^{n+3}$. Consequently, interactions with $n>0$ receive an addtional prominent enhancement from the large pericenter velocity of S301. Fast S-stars may therefore be particularly valuable for probing velocity-enhanced interactions. The same strategy can also be extended to other fast stars around supermassive black holes. A combined analysis of multiple stars would be helpful to break DM-astrophysics degeneracies, and open a broader avenue for probing dark matter interactions in high-density and high-velocity astrophysical environments.

\medskip
{\bf Acknowledgements.}~~\\
The authors acknowledge support from the National Natural Science Foundation of China (12441504).

\bibliography{refs}

\begin{thebibliography}{52}%
\makeatletter
\providecommand \@ifxundefined [1]{%
 \@ifx{#1\undefined}
}%
\providecommand \@ifnum [1]{%
 \ifnum #1\expandafter \@firstoftwo
 \else \expandafter \@secondoftwo
 \fi
}%
\providecommand \@ifx [1]{%
 \ifx #1\expandafter \@firstoftwo
 \else \expandafter \@secondoftwo
 \fi
}%
\providecommand \natexlab [1]{#1}%
\providecommand \enquote  [1]{``#1''}%
\providecommand \bibnamefont  [1]{#1}%
\providecommand \bibfnamefont [1]{#1}%
\providecommand \citenamefont [1]{#1}%
\providecommand \href@noop [0]{\@secondoftwo}%
\providecommand \href [0]{\begingroup \@sanitize@url \@href}%
\providecommand \@href[1]{\@@startlink{#1}\@@href}%
\providecommand \@@href[1]{\endgroup#1\@@endlink}%
\providecommand \@sanitize@url [0]{\catcode `\\12\catcode `\$12\catcode
  `\&12\catcode `\#12\catcode `\^12\catcode `\_12\catcode `\%12\relax}%
\providecommand \@@startlink[1]{}%
\providecommand \@@endlink[0]{}%
\providecommand \url  [0]{\begingroup\@sanitize@url \@url }%
\providecommand \@url [1]{\endgroup\@href {#1}{\urlprefix }}%
\providecommand \urlprefix  [0]{URL }%
\providecommand \Eprint [0]{\href }%
\providecommand \doibase [0]{http://dx.doi.org/}%
\providecommand \selectlanguage [0]{\@gobble}%
\providecommand \bibinfo  [0]{\@secondoftwo}%
\providecommand \bibfield  [0]{\@secondoftwo}%
\providecommand \translation [1]{[#1]}%
\providecommand \BibitemOpen [0]{}%
\providecommand \bibitemStop [0]{}%
\providecommand \bibitemNoStop [0]{.\EOS\space}%
\providecommand \EOS [0]{\spacefactor3000\relax}%
\providecommand \BibitemShut  [1]{\csname bibitem#1\endcsname}%
\let\auto@bib@innerbib\@empty
%</preamble>
\bibitem [{\citenamefont {Aghanim}\ \emph {et~al.}(2020)\citenamefont {Aghanim}
  \emph {et~al.}}]{Planck:2018vyg}%
  \BibitemOpen
  \bibfield  {author} {\bibinfo {author} {\bibfnamefont {N.}~\bibnamefont
  {Aghanim}} \emph {et~al.} (\bibinfo {collaboration} {Planck}),\ }\bibfield
  {title} {\enquote {\bibinfo {title} {{Planck 2018 results. VI. Cosmological
  parameters}},}\ }\href {\doibase 10.1051/0004-6361/201833910} {\bibfield
  {journal} {\bibinfo  {journal} {Astron. Astrophys.}\ }\textbf {\bibinfo
  {volume} {641}},\ \bibinfo {pages} {A6} (\bibinfo {year} {2020})},\ \bibinfo
  {note} {[Erratum: Astron.Astrophys. 652, C4 (2021)]},\ \Eprint
  {http://arxiv.org/abs/1807.06209} {arXiv:1807.06209 [astro-ph.CO]}
  \BibitemShut {NoStop}%
\bibitem [{\citenamefont {Arbey}\ and\ \citenamefont
  {Mahmoudi}(2021)}]{Arbey:2021gdg}%
  \BibitemOpen
  \bibfield  {author} {\bibinfo {author} {\bibfnamefont {A.}~\bibnamefont
  {Arbey}}\ and\ \bibinfo {author} {\bibfnamefont {F.}~\bibnamefont
  {Mahmoudi}},\ }\bibfield  {title} {\enquote {\bibinfo {title} {{Dark matter
  and the early Universe: a review}},}\ }\href {\doibase
  10.1016/j.ppnp.2021.103865} {\bibfield  {journal} {\bibinfo  {journal} {Prog.
  Part. Nucl. Phys.}\ }\textbf {\bibinfo {volume} {119}},\ \bibinfo {pages}
  {103865} (\bibinfo {year} {2021})},\ \Eprint
  {http://arxiv.org/abs/2104.11488} {arXiv:2104.11488 [hep-ph]} \BibitemShut
  {NoStop}%
\bibitem [{\citenamefont {Aalbers}\ \emph
  {et~al.}(2025{\natexlab{a}})\citenamefont {Aalbers} \emph
  {et~al.}}]{LZ:2024zvo}%
  \BibitemOpen
  \bibfield  {author} {\bibinfo {author} {\bibfnamefont {J.}~\bibnamefont
  {Aalbers}} \emph {et~al.} (\bibinfo {collaboration} {LZ}),\ }\bibfield
  {title} {\enquote {\bibinfo {title} {{Dark Matter Search Results from
  4.2{\,}{\,}Tonne-Years of Exposure of the LUX-ZEPLIN (LZ) Experiment}},}\
  }\href {\doibase 10.1103/4dyc-z8zf} {\bibfield  {journal} {\bibinfo
  {journal} {Phys. Rev. Lett.}\ }\textbf {\bibinfo {volume} {135}},\ \bibinfo
  {pages} {011802} (\bibinfo {year} {2025}{\natexlab{a}})},\ \Eprint
  {http://arxiv.org/abs/2410.17036} {arXiv:2410.17036 [hep-ex]} \BibitemShut
  {NoStop}%
\bibitem [{\citenamefont {Bo}\ \emph {et~al.}(2025)\citenamefont {Bo} \emph
  {et~al.}}]{PandaX:2024qfu}%
  \BibitemOpen
  \bibfield  {author} {\bibinfo {author} {\bibfnamefont {Zihao}\ \bibnamefont
  {Bo}} \emph {et~al.} (\bibinfo {collaboration} {PandaX}),\ }\bibfield
  {title} {\enquote {\bibinfo {title} {{Dark Matter Search Results from
  1.54{\,}{\,}Tonne{\textperiodcentered}Year Exposure of PandaX-4T}},}\ }\href
  {\doibase 10.1103/PhysRevLett.134.011805} {\bibfield  {journal} {\bibinfo
  {journal} {Phys. Rev. Lett.}\ }\textbf {\bibinfo {volume} {134}},\ \bibinfo
  {pages} {011805} (\bibinfo {year} {2025})},\ \Eprint
  {http://arxiv.org/abs/2408.00664} {arXiv:2408.00664 [hep-ex]} \BibitemShut
  {NoStop}%
\bibitem [{\citenamefont {Aprile}\ \emph {et~al.}(2023)\citenamefont {Aprile}
  \emph {et~al.}}]{XENON:2023cxc}%
  \BibitemOpen
  \bibfield  {author} {\bibinfo {author} {\bibfnamefont {E.}~\bibnamefont
  {Aprile}} \emph {et~al.} (\bibinfo {collaboration} {XENON}),\ }\bibfield
  {title} {\enquote {\bibinfo {title} {{First Dark Matter Search with Nuclear
  Recoils from the XENONnT Experiment}},}\ }\href {\doibase
  10.1103/PhysRevLett.131.041003} {\bibfield  {journal} {\bibinfo  {journal}
  {Phys. Rev. Lett.}\ }\textbf {\bibinfo {volume} {131}},\ \bibinfo {pages}
  {041003} (\bibinfo {year} {2023})},\ \Eprint
  {http://arxiv.org/abs/2303.14729} {arXiv:2303.14729 [hep-ex]} \BibitemShut
  {NoStop}%
\bibitem [{\citenamefont {Aprile}\ \emph {et~al.}(2018)\citenamefont {Aprile}
  \emph {et~al.}}]{XENON:2018voc}%
  \BibitemOpen
  \bibfield  {author} {\bibinfo {author} {\bibfnamefont {E.}~\bibnamefont
  {Aprile}} \emph {et~al.} (\bibinfo {collaboration} {XENON}),\ }\bibfield
  {title} {\enquote {\bibinfo {title} {{Dark Matter Search Results from a One
  Ton-Year Exposure of XENON1T}},}\ }\href {\doibase
  10.1103/PhysRevLett.121.111302} {\bibfield  {journal} {\bibinfo  {journal}
  {Phys. Rev. Lett.}\ }\textbf {\bibinfo {volume} {121}},\ \bibinfo {pages}
  {111302} (\bibinfo {year} {2018})},\ \Eprint
  {http://arxiv.org/abs/1805.12562} {arXiv:1805.12562 [astro-ph.CO]}
  \BibitemShut {NoStop}%
\bibitem [{\citenamefont {Akerib}\ \emph {et~al.}(2017)\citenamefont {Akerib}
  \emph {et~al.}}]{LUX:2016ggv}%
  \BibitemOpen
  \bibfield  {author} {\bibinfo {author} {\bibfnamefont {D.~S.}\ \bibnamefont
  {Akerib}} \emph {et~al.} (\bibinfo {collaboration} {LUX}),\ }\bibfield
  {title} {\enquote {\bibinfo {title} {{Results from a search for dark matter
  in the complete LUX exposure}},}\ }\href {\doibase
  10.1103/PhysRevLett.118.021303} {\bibfield  {journal} {\bibinfo  {journal}
  {Phys. Rev. Lett.}\ }\textbf {\bibinfo {volume} {118}},\ \bibinfo {pages}
  {021303} (\bibinfo {year} {2017})},\ \Eprint
  {http://arxiv.org/abs/1608.07648} {arXiv:1608.07648 [astro-ph.CO]}
  \BibitemShut {NoStop}%
\bibitem [{\citenamefont {Ajaj}\ \emph {et~al.}(2019)\citenamefont {Ajaj} \emph
  {et~al.}}]{DEAP:2019yzn}%
  \BibitemOpen
  \bibfield  {author} {\bibinfo {author} {\bibfnamefont {R.}~\bibnamefont
  {Ajaj}} \emph {et~al.} (\bibinfo {collaboration} {DEAP}),\ }\bibfield
  {title} {\enquote {\bibinfo {title} {{Search for dark matter with a 231-day
  exposure of liquid argon using DEAP-3600 at SNOLAB}},}\ }\href {\doibase
  10.1103/PhysRevD.100.022004} {\bibfield  {journal} {\bibinfo  {journal}
  {Phys. Rev. D}\ }\textbf {\bibinfo {volume} {100}},\ \bibinfo {pages}
  {022004} (\bibinfo {year} {2019})},\ \Eprint
  {http://arxiv.org/abs/1902.04048} {arXiv:1902.04048 [astro-ph.CO]}
  \BibitemShut {NoStop}%
\bibitem [{\citenamefont {Amole}\ \emph {et~al.}(2019)\citenamefont {Amole}
  \emph {et~al.}}]{PICO:2019vsc}%
  \BibitemOpen
  \bibfield  {author} {\bibinfo {author} {\bibfnamefont {C.}~\bibnamefont
  {Amole}} \emph {et~al.} (\bibinfo {collaboration} {PICO}),\ }\bibfield
  {title} {\enquote {\bibinfo {title} {{Dark Matter Search Results from the
  Complete Exposure of the PICO-60 C$_3$F$_8$ Bubble Chamber}},}\ }\href
  {\doibase 10.1103/PhysRevD.100.022001} {\bibfield  {journal} {\bibinfo
  {journal} {Phys. Rev. D}\ }\textbf {\bibinfo {volume} {100}},\ \bibinfo
  {pages} {022001} (\bibinfo {year} {2019})},\ \Eprint
  {http://arxiv.org/abs/1902.04031} {arXiv:1902.04031 [astro-ph.CO]}
  \BibitemShut {NoStop}%
\bibitem [{\citenamefont {Agnes}\ \emph {et~al.}(2018)\citenamefont {Agnes}
  \emph {et~al.}}]{DarkSide:2018bpj}%
  \BibitemOpen
  \bibfield  {author} {\bibinfo {author} {\bibfnamefont {P.}~\bibnamefont
  {Agnes}} \emph {et~al.} (\bibinfo {collaboration} {DarkSide}),\ }\bibfield
  {title} {\enquote {\bibinfo {title} {{Low-Mass Dark Matter Search with the
  DarkSide-50 Experiment}},}\ }\href {\doibase 10.1103/PhysRevLett.121.081307}
  {\bibfield  {journal} {\bibinfo  {journal} {Phys. Rev. Lett.}\ }\textbf
  {\bibinfo {volume} {121}},\ \bibinfo {pages} {081307} (\bibinfo {year}
  {2018})},\ \Eprint {http://arxiv.org/abs/1802.06994} {arXiv:1802.06994
  [astro-ph.HE]} \BibitemShut {NoStop}%
\bibitem [{\citenamefont {Baryakhtar}\ \emph {et~al.}(2017)\citenamefont
  {Baryakhtar}, \citenamefont {Bramante}, \citenamefont {Li}, \citenamefont
  {Linden},\ and\ \citenamefont {Raj}}]{Baryakhtar:2017dbj}%
  \BibitemOpen
  \bibfield  {author} {\bibinfo {author} {\bibfnamefont {Masha}\ \bibnamefont
  {Baryakhtar}}, \bibinfo {author} {\bibfnamefont {Joseph}\ \bibnamefont
  {Bramante}}, \bibinfo {author} {\bibfnamefont {Shirley~Weishi}\ \bibnamefont
  {Li}}, \bibinfo {author} {\bibfnamefont {Tim}\ \bibnamefont {Linden}}, \ and\
  \bibinfo {author} {\bibfnamefont {Nirmal}\ \bibnamefont {Raj}},\ }\bibfield
  {title} {\enquote {\bibinfo {title} {{Dark Kinetic Heating of Neutron Stars
  and An Infrared Window On WIMPs, SIMPs, and Pure Higgsinos}},}\ }\href
  {\doibase 10.1103/PhysRevLett.119.131801} {\bibfield  {journal} {\bibinfo
  {journal} {Phys. Rev. Lett.}\ }\textbf {\bibinfo {volume} {119}},\ \bibinfo
  {pages} {131801} (\bibinfo {year} {2017})},\ \Eprint
  {http://arxiv.org/abs/1704.01577} {arXiv:1704.01577 [hep-ph]} \BibitemShut
  {NoStop}%
\bibitem [{\citenamefont {Graham}\ \emph {et~al.}(2018)\citenamefont {Graham},
  \citenamefont {Janish}, \citenamefont {Narayan}, \citenamefont {Rajendran},\
  and\ \citenamefont {Riggins}}]{Graham:2018efk}%
  \BibitemOpen
  \bibfield  {author} {\bibinfo {author} {\bibfnamefont {Peter~W.}\
  \bibnamefont {Graham}}, \bibinfo {author} {\bibfnamefont {Ryan}\ \bibnamefont
  {Janish}}, \bibinfo {author} {\bibfnamefont {Vijay}\ \bibnamefont {Narayan}},
  \bibinfo {author} {\bibfnamefont {Surjeet}\ \bibnamefont {Rajendran}}, \ and\
  \bibinfo {author} {\bibfnamefont {Paul}\ \bibnamefont {Riggins}},\ }\bibfield
   {title} {\enquote {\bibinfo {title} {{White Dwarfs as Dark Matter
  Detectors}},}\ }\href {\doibase 10.1103/PhysRevD.98.115027} {\bibfield
  {journal} {\bibinfo  {journal} {Phys. Rev. D}\ }\textbf {\bibinfo {volume}
  {98}},\ \bibinfo {pages} {115027} (\bibinfo {year} {2018})},\ \Eprint
  {http://arxiv.org/abs/1805.07381} {arXiv:1805.07381 [hep-ph]} \BibitemShut
  {NoStop}%
\bibitem [{\citenamefont {Leane}\ and\ \citenamefont
  {Smirnov}(2021)}]{Leane:2020wob}%
  \BibitemOpen
  \bibfield  {author} {\bibinfo {author} {\bibfnamefont {Rebecca~K.}\
  \bibnamefont {Leane}}\ and\ \bibinfo {author} {\bibfnamefont {Juri}\
  \bibnamefont {Smirnov}},\ }\bibfield  {title} {\enquote {\bibinfo {title}
  {{Exoplanets as Sub-GeV Dark Matter Detectors}},}\ }\href {\doibase
  10.1103/PhysRevLett.126.161101} {\bibfield  {journal} {\bibinfo  {journal}
  {Phys. Rev. Lett.}\ }\textbf {\bibinfo {volume} {126}},\ \bibinfo {pages}
  {161101} (\bibinfo {year} {2021})},\ \Eprint
  {http://arxiv.org/abs/2010.00015} {arXiv:2010.00015 [hep-ph]} \BibitemShut
  {NoStop}%
\bibitem [{\citenamefont {Wadekar}\ and\ \citenamefont
  {Farrar}(2021)}]{Wadekar:2019mpc}%
  \BibitemOpen
  \bibfield  {author} {\bibinfo {author} {\bibfnamefont {Digvijay}\
  \bibnamefont {Wadekar}}\ and\ \bibinfo {author} {\bibfnamefont {Glennys~R.}\
  \bibnamefont {Farrar}},\ }\bibfield  {title} {\enquote {\bibinfo {title}
  {{Gas-rich dwarf galaxies as a new probe of dark matter interactions with
  ordinary matter}},}\ }\href {\doibase 10.1103/PhysRevD.103.123028} {\bibfield
   {journal} {\bibinfo  {journal} {Phys. Rev. D}\ }\textbf {\bibinfo {volume}
  {103}},\ \bibinfo {pages} {123028} (\bibinfo {year} {2021})},\ \Eprint
  {http://arxiv.org/abs/1903.12190} {arXiv:1903.12190 [hep-ph]} \BibitemShut
  {NoStop}%
\bibitem [{\citenamefont {Bringmann}\ and\ \citenamefont
  {Pospelov}(2019)}]{Bringmann:2018cvk}%
  \BibitemOpen
  \bibfield  {author} {\bibinfo {author} {\bibfnamefont {Torsten}\ \bibnamefont
  {Bringmann}}\ and\ \bibinfo {author} {\bibfnamefont {Maxim}\ \bibnamefont
  {Pospelov}},\ }\bibfield  {title} {\enquote {\bibinfo {title} {{Novel direct
  detection constraints on light dark matter}},}\ }\href {\doibase
  10.1103/PhysRevLett.122.171801} {\bibfield  {journal} {\bibinfo  {journal}
  {Phys. Rev. Lett.}\ }\textbf {\bibinfo {volume} {122}},\ \bibinfo {pages}
  {171801} (\bibinfo {year} {2019})},\ \Eprint
  {http://arxiv.org/abs/1810.10543} {arXiv:1810.10543 [hep-ph]} \BibitemShut
  {NoStop}%
\bibitem [{\citenamefont {Dvorkin}\ \emph {et~al.}(2014)\citenamefont
  {Dvorkin}, \citenamefont {Blum},\ and\ \citenamefont
  {Kamionkowski}}]{Dvorkin:2013cea}%
  \BibitemOpen
  \bibfield  {author} {\bibinfo {author} {\bibfnamefont {Cora}\ \bibnamefont
  {Dvorkin}}, \bibinfo {author} {\bibfnamefont {Kfir}\ \bibnamefont {Blum}}, \
  and\ \bibinfo {author} {\bibfnamefont {Marc}\ \bibnamefont {Kamionkowski}},\
  }\bibfield  {title} {\enquote {\bibinfo {title} {{Constraining Dark
  Matter-Baryon Scattering with Linear Cosmology}},}\ }\href {\doibase
  10.1103/PhysRevD.89.023519} {\bibfield  {journal} {\bibinfo  {journal} {Phys.
  Rev. D}\ }\textbf {\bibinfo {volume} {89}},\ \bibinfo {pages} {023519}
  (\bibinfo {year} {2014})},\ \Eprint {http://arxiv.org/abs/1311.2937}
  {arXiv:1311.2937 [astro-ph.CO]} \BibitemShut {NoStop}%
\bibitem [{\citenamefont {Abbasi}\ \emph {et~al.}(2022)\citenamefont {Abbasi}
  \emph {et~al.}}]{IceCube:2021xzo}%
  \BibitemOpen
  \bibfield  {author} {\bibinfo {author} {\bibfnamefont {R.}~\bibnamefont
  {Abbasi}} \emph {et~al.} (\bibinfo {collaboration} {IceCube}),\ }\bibfield
  {title} {\enquote {\bibinfo {title} {{Search for GeV-scale dark matter
  annihilation in the Sun with IceCube DeepCore}},}\ }\href {\doibase
  10.1103/PhysRevD.105.062004} {\bibfield  {journal} {\bibinfo  {journal}
  {Phys. Rev. D}\ }\textbf {\bibinfo {volume} {105}},\ \bibinfo {pages}
  {062004} (\bibinfo {year} {2022})},\ \Eprint
  {http://arxiv.org/abs/2111.09970} {arXiv:2111.09970 [astro-ph.HE]}
  \BibitemShut {NoStop}%
\bibitem [{\citenamefont {Lin}\ \emph {et~al.}(2025)\citenamefont {Lin},
  \citenamefont {Lu},\ and\ \citenamefont {Song}}]{Lin:2025mez}%
  \BibitemOpen
  \bibfield  {author} {\bibinfo {author} {\bibfnamefont {Yugen}\ \bibnamefont
  {Lin}}, \bibinfo {author} {\bibfnamefont {Chih-Ting}\ \bibnamefont {Lu}}, \
  and\ \bibinfo {author} {\bibfnamefont {Ningqiang}\ \bibnamefont {Song}},\
  }\bibfield  {title} {\enquote {\bibinfo {title} {{Supernova cooling from
  neutrino-devouring dark matter}},}\ }\href@noop {} {\  (\bibinfo {year}
  {2025})},\ \Eprint {http://arxiv.org/abs/2507.22124} {arXiv:2507.22124
  [hep-ph]} \BibitemShut {NoStop}%
\bibitem [{\citenamefont {Bi}\ \emph {et~al.}(2023)\citenamefont {Bi},
  \citenamefont {Gao}, \citenamefont {Jin}, \citenamefont {Lin},\ and\
  \citenamefont {Xiang}}]{Bi:2021njb}%
  \BibitemOpen
  \bibfield  {author} {\bibinfo {author} {\bibfnamefont {Xiao-jun}\
  \bibnamefont {Bi}}, \bibinfo {author} {\bibfnamefont {Yu}~\bibnamefont
  {Gao}}, \bibinfo {author} {\bibfnamefont {Mingjie}\ \bibnamefont {Jin}},
  \bibinfo {author} {\bibfnamefont {Yugen}\ \bibnamefont {Lin}}, \ and\
  \bibinfo {author} {\bibfnamefont {Qian-Fei}\ \bibnamefont {Xiang}},\
  }\bibfield  {title} {\enquote {\bibinfo {title} {{Soft scattering evaporation
  of dark matter subhalos by inner galactic gases}},}\ }\href {\doibase
  10.1140/epjc/s10052-023-11987-w} {\bibfield  {journal} {\bibinfo  {journal}
  {Eur. Phys. J. C}\ }\textbf {\bibinfo {volume} {83}},\ \bibinfo {pages} {808}
  (\bibinfo {year} {2023})},\ \Eprint {http://arxiv.org/abs/2112.01260}
  {arXiv:2112.01260 [astro-ph.CO]} \BibitemShut {NoStop}%
\bibitem [{\citenamefont {Meighen-Berger}\ \emph {et~al.}(2026)\citenamefont
  {Meighen-Berger}, \citenamefont {Gustafson}, \citenamefont {Bell},
  \citenamefont {Newstead}, \citenamefont {Robles},\ and\ \citenamefont
  {Shoemaker}}]{Meighen-Berger:2026idy}%
  \BibitemOpen
  \bibfield  {author} {\bibinfo {author} {\bibfnamefont {Stephan~A.}\
  \bibnamefont {Meighen-Berger}}, \bibinfo {author} {\bibfnamefont {R.~Andrew}\
  \bibnamefont {Gustafson}}, \bibinfo {author} {\bibfnamefont {Nicole~F.}\
  \bibnamefont {Bell}}, \bibinfo {author} {\bibfnamefont {Jayden~L.}\
  \bibnamefont {Newstead}}, \bibinfo {author} {\bibfnamefont {Sandra}\
  \bibnamefont {Robles}}, \ and\ \bibinfo {author} {\bibfnamefont {Ian~M.}\
  \bibnamefont {Shoemaker}},\ }\bibfield  {title} {\enquote {\bibinfo {title}
  {{Dark matter energy exchange in stars orbiting supermassive black holes}},}\
  }\href@noop {} {\  (\bibinfo {year} {2026})},\ \Eprint
  {http://arxiv.org/abs/2607.00840} {arXiv:2607.00840 [hep-ph]} \BibitemShut
  {NoStop}%
\bibitem [{\citenamefont {Schodel}\ \emph {et~al.}(2002)\citenamefont {Schodel}
  \emph {et~al.}}]{Schodel:2002py}%
  \BibitemOpen
  \bibfield  {author} {\bibinfo {author} {\bibfnamefont {R.}~\bibnamefont
  {Schodel}} \emph {et~al.},\ }\bibfield  {title} {\enquote {\bibinfo {title}
  {{A Star in a 15.2 year orbit around the supermassive black hole at the
  center of the Milky Way}},}\ }\href {\doibase 10.1038/nature01121} {\bibfield
   {journal} {\bibinfo  {journal} {Nature}\ }\textbf {\bibinfo {volume}
  {419}},\ \bibinfo {pages} {694--696} (\bibinfo {year} {2002})},\ \Eprint
  {http://arxiv.org/abs/astro-ph/0210426} {arXiv:astro-ph/0210426} \BibitemShut
  {NoStop}%
\bibitem [{\citenamefont {Abuter}\ \emph {et~al.}(2019)\citenamefont {Abuter}
  \emph {et~al.}}]{Gravity:2019nxk}%
  \BibitemOpen
  \bibfield  {author} {\bibinfo {author} {\bibfnamefont {R.}~\bibnamefont
  {Abuter}} \emph {et~al.} (\bibinfo {collaboration} {Gravity}),\ }\bibfield
  {title} {\enquote {\bibinfo {title} {{A geometric distance measurement to the
  Galactic center black hole with 0.3{\%} uncertainty}},}\ }\href {\doibase
  10.1051/0004-6361/201935656} {\bibfield  {journal} {\bibinfo  {journal}
  {Astron. Astrophys.}\ }\textbf {\bibinfo {volume} {625}} (\bibinfo {year}
  {2019}),\ 10.1051/0004-6361/201935656},\ \Eprint
  {http://arxiv.org/abs/1904.05721} {arXiv:1904.05721 [astro-ph.GA]}
  \BibitemShut {NoStop}%
\bibitem [{\citenamefont {Abuter}\ \emph {et~al.}(2018)\citenamefont {Abuter}
  \emph {et~al.}}]{GRAVITY:2018ofz}%
  \BibitemOpen
  \bibfield  {author} {\bibinfo {author} {\bibfnamefont {R.}~\bibnamefont
  {Abuter}} \emph {et~al.} (\bibinfo {collaboration} {GRAVITY}),\ }\bibfield
  {title} {\enquote {\bibinfo {title} {{Detection of the gravitational redshift
  in the orbit of the star S2 near the Galactic centre massive black hole}},}\
  }\href {\doibase 10.1051/0004-6361/201833718} {\bibfield  {journal} {\bibinfo
   {journal} {Astron. Astrophys.}\ }\textbf {\bibinfo {volume} {615}},\
  \bibinfo {pages} {L15} (\bibinfo {year} {2018})},\ \Eprint
  {http://arxiv.org/abs/1807.09409} {arXiv:1807.09409 [astro-ph.GA]}
  \BibitemShut {NoStop}%
\bibitem [{\citenamefont {Abuter}\ \emph {et~al.}(2020)\citenamefont {Abuter}
  \emph {et~al.}}]{GRAVITY:2020gka}%
  \BibitemOpen
  \bibfield  {author} {\bibinfo {author} {\bibfnamefont {R.}~\bibnamefont
  {Abuter}} \emph {et~al.} (\bibinfo {collaboration} {GRAVITY}),\ }\bibfield
  {title} {\enquote {\bibinfo {title} {{Detection of the Schwarzschild
  precession in the orbit of the star S2 near the Galactic centre massive black
  hole}},}\ }\href {\doibase 10.1051/0004-6361/202037813} {\bibfield  {journal}
  {\bibinfo  {journal} {Astron. Astrophys.}\ }\textbf {\bibinfo {volume}
  {636}},\ \bibinfo {pages} {L5} (\bibinfo {year} {2020})},\ \Eprint
  {http://arxiv.org/abs/2004.07187} {arXiv:2004.07187 [astro-ph.GA]}
  \BibitemShut {NoStop}%
\bibitem [{\citenamefont {Hees}\ \emph {et~al.}(2017)\citenamefont {Hees} \emph
  {et~al.}}]{Hees:2017aal}%
  \BibitemOpen
  \bibfield  {author} {\bibinfo {author} {\bibfnamefont {A.}~\bibnamefont
  {Hees}} \emph {et~al.},\ }\bibfield  {title} {\enquote {\bibinfo {title}
  {{Testing General Relativity with stellar orbits around the supermassive
  black hole in our Galactic center}},}\ }\href {\doibase
  10.1103/PhysRevLett.118.211101} {\bibfield  {journal} {\bibinfo  {journal}
  {Phys. Rev. Lett.}\ }\textbf {\bibinfo {volume} {118}},\ \bibinfo {pages}
  {211101} (\bibinfo {year} {2017})},\ \Eprint
  {http://arxiv.org/abs/1705.07902} {arXiv:1705.07902 [astro-ph.GA]}
  \BibitemShut {NoStop}%
\bibitem [{\citenamefont {Abuter}\ \emph {et~al.}(2022)\citenamefont {Abuter}
  \emph {et~al.}}]{GRAVITY:2021xju}%
  \BibitemOpen
  \bibfield  {author} {\bibinfo {author} {\bibfnamefont {R.}~\bibnamefont
  {Abuter}} \emph {et~al.} (\bibinfo {collaboration} {GRAVITY}),\ }\bibfield
  {title} {\enquote {\bibinfo {title} {{Mass distribution in the Galactic
  Center based on interferometric astrometry of multiple stellar orbits}},}\
  }\href {\doibase 10.1051/0004-6361/202142465} {\bibfield  {journal} {\bibinfo
   {journal} {Astron. Astrophys.}\ }\textbf {\bibinfo {volume} {657}},\
  \bibinfo {pages} {L12} (\bibinfo {year} {2022})},\ \Eprint
  {http://arxiv.org/abs/2112.07478} {arXiv:2112.07478 [astro-ph.GA]}
  \BibitemShut {NoStop}%
\bibitem [{\citenamefont {Dayem}\ \emph {et~al.}(2026)\citenamefont {Dayem}
  \emph {et~al.}}]{Dayem:2026ktt}%
  \BibitemOpen
  \bibfield  {author} {\bibinfo {author} {\bibfnamefont {K.~Abd~El}\
  \bibnamefont {Dayem}} \emph {et~al.},\ }\bibfield  {title} {\enquote
  {\bibinfo {title} {{Discovery of a star sensitive to the spin of Sgr A*}},}\
  }\href@noop {} {\  (\bibinfo {year} {2026})},\ \Eprint
  {http://arxiv.org/abs/2607.12664} {arXiv:2607.12664 [astro-ph.GA]}
  \BibitemShut {NoStop}%
\bibitem [{\citenamefont {Quinlan}\ \emph {et~al.}(1995)\citenamefont
  {Quinlan}, \citenamefont {Hernquist},\ and\ \citenamefont
  {Sigurdsson}}]{Quinlan:1994ed}%
  \BibitemOpen
  \bibfield  {author} {\bibinfo {author} {\bibfnamefont {Gerald~D.}\
  \bibnamefont {Quinlan}}, \bibinfo {author} {\bibfnamefont {Lars}\
  \bibnamefont {Hernquist}}, \ and\ \bibinfo {author} {\bibfnamefont {Steinn}\
  \bibnamefont {Sigurdsson}},\ }\bibfield  {title} {\enquote {\bibinfo {title}
  {{Models of Galaxies with Central Black Holes: Adiabatic Growth in Spherical
  Galaxies}},}\ }\href {\doibase 10.1086/175295} {\bibfield  {journal}
  {\bibinfo  {journal} {Astrophys. J.}\ }\textbf {\bibinfo {volume} {440}},\
  \bibinfo {pages} {554--564} (\bibinfo {year} {1995})},\ \Eprint
  {http://arxiv.org/abs/astro-ph/9407005} {arXiv:astro-ph/9407005} \BibitemShut
  {NoStop}%
\bibitem [{\citenamefont {Sadeghian}\ \emph {et~al.}(2013)\citenamefont
  {Sadeghian}, \citenamefont {Ferrer},\ and\ \citenamefont
  {Will}}]{Sadeghian:2013laa}%
  \BibitemOpen
  \bibfield  {author} {\bibinfo {author} {\bibfnamefont {Laleh}\ \bibnamefont
  {Sadeghian}}, \bibinfo {author} {\bibfnamefont {Francesc}\ \bibnamefont
  {Ferrer}}, \ and\ \bibinfo {author} {\bibfnamefont {Clifford~M.}\
  \bibnamefont {Will}},\ }\bibfield  {title} {\enquote {\bibinfo {title} {{Dark
  matter distributions around massive black holes: A general relativistic
  analysis}},}\ }\href {\doibase 10.1103/PhysRevD.88.063522} {\bibfield
  {journal} {\bibinfo  {journal} {Phys. Rev. D}\ }\textbf {\bibinfo {volume}
  {88}},\ \bibinfo {pages} {063522} (\bibinfo {year} {2013})},\ \Eprint
  {http://arxiv.org/abs/1305.2619} {arXiv:1305.2619 [astro-ph.GA]} \BibitemShut
  {NoStop}%
\bibitem [{\citenamefont {Navarro}\ \emph {et~al.}(1996)\citenamefont
  {Navarro}, \citenamefont {Frenk},\ and\ \citenamefont
  {White}}]{Navarro:1995iw}%
  \BibitemOpen
  \bibfield  {author} {\bibinfo {author} {\bibfnamefont {Julio~F.}\
  \bibnamefont {Navarro}}, \bibinfo {author} {\bibfnamefont {Carlos~S.}\
  \bibnamefont {Frenk}}, \ and\ \bibinfo {author} {\bibfnamefont {Simon D.~M.}\
  \bibnamefont {White}},\ }\bibfield  {title} {\enquote {\bibinfo {title} {{The
  Structure of cold dark matter halos}},}\ }\href {\doibase 10.1086/177173}
  {\bibfield  {journal} {\bibinfo  {journal} {Astrophys. J.}\ }\textbf
  {\bibinfo {volume} {462}},\ \bibinfo {pages} {563--575} (\bibinfo {year}
  {1996})},\ \Eprint {http://arxiv.org/abs/astro-ph/9508025}
  {arXiv:astro-ph/9508025} \BibitemShut {NoStop}%
\bibitem [{\citenamefont {Cirelli}\ \emph {et~al.}(2011)\citenamefont
  {Cirelli}, \citenamefont {Corcella}, \citenamefont {Hektor}, \citenamefont
  {Hutsi}, \citenamefont {Kadastik}, \citenamefont {Panci}, \citenamefont
  {Raidal}, \citenamefont {Sala},\ and\ \citenamefont
  {Strumia}}]{Cirelli:2010xx}%
  \BibitemOpen
  \bibfield  {author} {\bibinfo {author} {\bibfnamefont {Marco}\ \bibnamefont
  {Cirelli}}, \bibinfo {author} {\bibfnamefont {Gennaro}\ \bibnamefont
  {Corcella}}, \bibinfo {author} {\bibfnamefont {Andi}\ \bibnamefont {Hektor}},
  \bibinfo {author} {\bibfnamefont {Gert}\ \bibnamefont {Hutsi}}, \bibinfo
  {author} {\bibfnamefont {Mario}\ \bibnamefont {Kadastik}}, \bibinfo {author}
  {\bibfnamefont {Paolo}\ \bibnamefont {Panci}}, \bibinfo {author}
  {\bibfnamefont {Martti}\ \bibnamefont {Raidal}}, \bibinfo {author}
  {\bibfnamefont {Filippo}\ \bibnamefont {Sala}}, \ and\ \bibinfo {author}
  {\bibfnamefont {Alessandro}\ \bibnamefont {Strumia}},\ }\bibfield  {title}
  {\enquote {\bibinfo {title} {{PPPC 4 DM ID: A Poor Particle Physicist
  Cookbook for Dark Matter Indirect Detection}},}\ }\href {\doibase
  10.1088/1475-7516/2012/10/E01} {\bibfield  {journal} {\bibinfo  {journal}
  {JCAP}\ }\textbf {\bibinfo {volume} {03}},\ \bibinfo {pages} {051} (\bibinfo
  {year} {2011})},\ \bibinfo {note} {[Erratum: JCAP 10, E01 (2012)]},\ \Eprint
  {http://arxiv.org/abs/1012.4515} {arXiv:1012.4515 [hep-ph]} \BibitemShut
  {NoStop}%
\bibitem [{\citenamefont {Gnedin}\ and\ \citenamefont
  {Primack}(2004)}]{Gnedin:2003rj}%
  \BibitemOpen
  \bibfield  {author} {\bibinfo {author} {\bibfnamefont {Oleg~Y.}\ \bibnamefont
  {Gnedin}}\ and\ \bibinfo {author} {\bibfnamefont {Joel~R.}\ \bibnamefont
  {Primack}},\ }\bibfield  {title} {\enquote {\bibinfo {title} {{Dark Matter
  Profile in the Galactic Center}},}\ }\href {\doibase
  10.1103/PhysRevLett.93.061302} {\bibfield  {journal} {\bibinfo  {journal}
  {Phys. Rev. Lett.}\ }\textbf {\bibinfo {volume} {93}},\ \bibinfo {pages}
  {061302} (\bibinfo {year} {2004})},\ \Eprint
  {http://arxiv.org/abs/astro-ph/0308385} {arXiv:astro-ph/0308385} \BibitemShut
  {NoStop}%
\bibitem [{\citenamefont {Iocco}\ and\ \citenamefont
  {Benito}(2017)}]{Iocco:2016itg}%
  \BibitemOpen
  \bibfield  {author} {\bibinfo {author} {\bibfnamefont {Fabio}\ \bibnamefont
  {Iocco}}\ and\ \bibinfo {author} {\bibfnamefont {Maria}\ \bibnamefont
  {Benito}},\ }\bibfield  {title} {\enquote {\bibinfo {title} {{An estimate of
  the DM profile in the Galactic bulge region}},}\ }\href {\doibase
  10.1016/j.dark.2016.12.004} {\bibfield  {journal} {\bibinfo  {journal} {Phys.
  Dark Univ.}\ }\textbf {\bibinfo {volume} {15}},\ \bibinfo {pages} {90--95}
  (\bibinfo {year} {2017})},\ \Eprint {http://arxiv.org/abs/1611.09861}
  {arXiv:1611.09861 [astro-ph.GA]} \BibitemShut {NoStop}%
\bibitem [{\citenamefont {Hooper}(2017)}]{Hooper:2016ggc}%
  \BibitemOpen
  \bibfield  {author} {\bibinfo {author} {\bibfnamefont {Dan}\ \bibnamefont
  {Hooper}},\ }\bibfield  {title} {\enquote {\bibinfo {title} {{The Density of
  Dark Matter in the Galactic Bulge and Implications for Indirect
  Detection}},}\ }\href {\doibase 10.1016/j.dark.2016.11.005} {\bibfield
  {journal} {\bibinfo  {journal} {Phys. Dark Univ.}\ }\textbf {\bibinfo
  {volume} {15}},\ \bibinfo {pages} {53--56} (\bibinfo {year} {2017})},\
  \Eprint {http://arxiv.org/abs/1608.00003} {arXiv:1608.00003 [astro-ph.HE]}
  \BibitemShut {NoStop}%
\bibitem [{\citenamefont {Baumgart}\ \emph {et~al.}(2026)\citenamefont
  {Baumgart}, \citenamefont {Bottaro}, \citenamefont {Redigolo}, \citenamefont
  {Rodd},\ and\ \citenamefont {Slatyer}}]{Baumgart:2025dov}%
  \BibitemOpen
  \bibfield  {author} {\bibinfo {author} {\bibfnamefont {Matthew}\ \bibnamefont
  {Baumgart}}, \bibinfo {author} {\bibfnamefont {Salvatore}\ \bibnamefont
  {Bottaro}}, \bibinfo {author} {\bibfnamefont {Diego}\ \bibnamefont
  {Redigolo}}, \bibinfo {author} {\bibfnamefont {Nicholas~L.}\ \bibnamefont
  {Rodd}}, \ and\ \bibinfo {author} {\bibfnamefont {Tracy~R.}\ \bibnamefont
  {Slatyer}},\ }\bibfield  {title} {\enquote {\bibinfo {title} {{Testing real
  WIMPs with CTAO}},}\ }\href {\doibase 10.1007/JHEP02(2026)213} {\bibfield
  {journal} {\bibinfo  {journal} {JHEP}\ }\textbf {\bibinfo {volume} {02}},\
  \bibinfo {pages} {213} (\bibinfo {year} {2026})},\ \Eprint
  {http://arxiv.org/abs/2507.15937} {arXiv:2507.15937 [hep-ph]} \BibitemShut
  {NoStop}%
\bibitem [{\citenamefont {Gondolo}\ and\ \citenamefont
  {Silk}(1999)}]{Gondolo:1999ef}%
  \BibitemOpen
  \bibfield  {author} {\bibinfo {author} {\bibfnamefont {Paolo}\ \bibnamefont
  {Gondolo}}\ and\ \bibinfo {author} {\bibfnamefont {Joseph}\ \bibnamefont
  {Silk}},\ }\bibfield  {title} {\enquote {\bibinfo {title} {{Dark matter
  annihilation at the galactic center}},}\ }\href {\doibase
  10.1103/PhysRevLett.83.1719} {\bibfield  {journal} {\bibinfo  {journal}
  {Phys. Rev. Lett.}\ }\textbf {\bibinfo {volume} {83}},\ \bibinfo {pages}
  {1719--1722} (\bibinfo {year} {1999})},\ \Eprint
  {http://arxiv.org/abs/astro-ph/9906391} {arXiv:astro-ph/9906391} \BibitemShut
  {NoStop}%
\bibitem [{\citenamefont {Aalbers}\ \emph
  {et~al.}(2025{\natexlab{b}})\citenamefont {Aalbers} \emph
  {et~al.}}]{LZ:2025iaw}%
  \BibitemOpen
  \bibfield  {author} {\bibinfo {author} {\bibfnamefont {J.}~\bibnamefont
  {Aalbers}} \emph {et~al.} (\bibinfo {collaboration} {LZ}),\ }\bibfield
  {title} {\enquote {\bibinfo {title} {{New Constraints on Cosmic Ray-Boosted
  Dark Matter from the LUX-ZEPLIN Experiment}},}\ }\href {\doibase
  10.1103/nr92-jvt3} {\bibfield  {journal} {\bibinfo  {journal} {Phys. Rev.
  Lett.}\ }\textbf {\bibinfo {volume} {134}},\ \bibinfo {pages} {241801}
  (\bibinfo {year} {2025}{\natexlab{b}})},\ \Eprint
  {http://arxiv.org/abs/2503.18158} {arXiv:2503.18158 [hep-ex]} \BibitemShut
  {NoStop}%
\bibitem [{\citenamefont {Kennedy}\ \emph {et~al.}(2014)\citenamefont
  {Kennedy}, \citenamefont {Frenk}, \citenamefont {Cole},\ and\ \citenamefont
  {Benson}}]{Kennedy:2013uta}%
  \BibitemOpen
  \bibfield  {author} {\bibinfo {author} {\bibfnamefont {Rachel}\ \bibnamefont
  {Kennedy}}, \bibinfo {author} {\bibfnamefont {Carlos}\ \bibnamefont {Frenk}},
  \bibinfo {author} {\bibfnamefont {Shaun}\ \bibnamefont {Cole}}, \ and\
  \bibinfo {author} {\bibfnamefont {Andrew}\ \bibnamefont {Benson}},\
  }\bibfield  {title} {\enquote {\bibinfo {title} {{Constraining the warm dark
  matter particle mass with Milky Way satellites}},}\ }\href {\doibase
  10.1093/mnras/stu719} {\bibfield  {journal} {\bibinfo  {journal} {Mon. Not.
  Roy. Astron. Soc.}\ }\textbf {\bibinfo {volume} {442}},\ \bibinfo {pages}
  {2487--2495} (\bibinfo {year} {2014})},\ \Eprint
  {http://arxiv.org/abs/1310.7739} {arXiv:1310.7739 [astro-ph.CO]} \BibitemShut
  {NoStop}%
\bibitem [{\citenamefont {Nadler}\ \emph {et~al.}(2021)\citenamefont {Nadler}
  \emph {et~al.}}]{DES:2020fxi}%
  \BibitemOpen
  \bibfield  {author} {\bibinfo {author} {\bibfnamefont {E.~O.}\ \bibnamefont
  {Nadler}} \emph {et~al.} (\bibinfo {collaboration} {DES}),\ }\bibfield
  {title} {\enquote {\bibinfo {title} {{Milky Way Satellite Census. III.
  Constraints on Dark Matter Properties from Observations of Milky Way
  Satellite Galaxies}},}\ }\href {\doibase 10.1103/PhysRevLett.126.091101}
  {\bibfield  {journal} {\bibinfo  {journal} {Phys. Rev. Lett.}\ }\textbf
  {\bibinfo {volume} {126}},\ \bibinfo {pages} {091101} (\bibinfo {year}
  {2021})},\ \Eprint {http://arxiv.org/abs/2008.00022} {arXiv:2008.00022
  [astro-ph.CO]} \BibitemShut {NoStop}%
\bibitem [{\citenamefont {Nadler}\ \emph {et~al.}(2019)\citenamefont {Nadler},
  \citenamefont {Gluscevic}, \citenamefont {Boddy},\ and\ \citenamefont
  {Wechsler}}]{Nadler:2019zrb}%
  \BibitemOpen
  \bibfield  {author} {\bibinfo {author} {\bibfnamefont {Ethan~O.}\
  \bibnamefont {Nadler}}, \bibinfo {author} {\bibfnamefont {Vera}\ \bibnamefont
  {Gluscevic}}, \bibinfo {author} {\bibfnamefont {Kimberly~K.}\ \bibnamefont
  {Boddy}}, \ and\ \bibinfo {author} {\bibfnamefont {Risa~H.}\ \bibnamefont
  {Wechsler}},\ }\bibfield  {title} {\enquote {\bibinfo {title} {{Constraints
  on Dark Matter Microphysics from the Milky Way Satellite Population}},}\
  }\href {\doibase 10.3847/2041-8213/ab1eb2} {\bibfield  {journal} {\bibinfo
  {journal} {Astrophys. J. Lett.}\ }\textbf {\bibinfo {volume} {878}},\
  \bibinfo {pages} {32} (\bibinfo {year} {2019})},\ \bibinfo {note} {[Erratum:
  Astrophys.J.Lett. 897, L46 (2020), Erratum: Astrophys.J. 897, L46 (2020)]},\
  \Eprint {http://arxiv.org/abs/1904.10000} {arXiv:1904.10000 [astro-ph.CO]}
  \BibitemShut {NoStop}%
\bibitem [{\citenamefont {Boehm}\ and\ \citenamefont
  {Schaeffer}(2005)}]{Boehm:2004th}%
  \BibitemOpen
  \bibfield  {author} {\bibinfo {author} {\bibfnamefont {Celine}\ \bibnamefont
  {Boehm}}\ and\ \bibinfo {author} {\bibfnamefont {Richard}\ \bibnamefont
  {Schaeffer}},\ }\bibfield  {title} {\enquote {\bibinfo {title} {{Constraints
  on dark matter interactions from structure formation: Damping lengths}},}\
  }\href {\doibase 10.1051/0004-6361:20042238} {\bibfield  {journal} {\bibinfo
  {journal} {Astron. Astrophys.}\ }\textbf {\bibinfo {volume} {438}},\ \bibinfo
  {pages} {419--442} (\bibinfo {year} {2005})},\ \Eprint
  {http://arxiv.org/abs/astro-ph/0410591} {arXiv:astro-ph/0410591} \BibitemShut
  {NoStop}%
\bibitem [{\citenamefont {Ali-Ha{\"\i}moud}\ \emph {et~al.}(2015)\citenamefont
  {Ali-Ha{\"\i}moud}, \citenamefont {Chluba},\ and\ \citenamefont
  {Kamionkowski}}]{Ali-Haimoud:2015pwa}%
  \BibitemOpen
  \bibfield  {author} {\bibinfo {author} {\bibfnamefont {Yacine}\ \bibnamefont
  {Ali-Ha{\"\i}moud}}, \bibinfo {author} {\bibfnamefont {Jens}\ \bibnamefont
  {Chluba}}, \ and\ \bibinfo {author} {\bibfnamefont {Marc}\ \bibnamefont
  {Kamionkowski}},\ }\bibfield  {title} {\enquote {\bibinfo {title}
  {{Constraints on Dark Matter Interactions with Standard Model Particles from
  Cosmic Microwave Background Spectral Distortions}},}\ }\href {\doibase
  10.1103/PhysRevLett.115.071304} {\bibfield  {journal} {\bibinfo  {journal}
  {Phys. Rev. Lett.}\ }\textbf {\bibinfo {volume} {115}},\ \bibinfo {pages}
  {071304} (\bibinfo {year} {2015})},\ \Eprint
  {http://arxiv.org/abs/1506.04745} {arXiv:1506.04745 [astro-ph.CO]}
  \BibitemShut {NoStop}%
\bibitem [{\citenamefont {Ali-Ha{\"\i}moud}(2021)}]{Ali-Haimoud:2021lka}%
  \BibitemOpen
  \bibfield  {author} {\bibinfo {author} {\bibfnamefont {Yacine}\ \bibnamefont
  {Ali-Ha{\"\i}moud}},\ }\bibfield  {title} {\enquote {\bibinfo {title}
  {{Testing dark matter interactions with CMB spectral distortions}},}\ }\href
  {\doibase 10.1103/PhysRevD.103.043541} {\bibfield  {journal} {\bibinfo
  {journal} {Phys. Rev. D}\ }\textbf {\bibinfo {volume} {103}},\ \bibinfo
  {pages} {043541} (\bibinfo {year} {2021})},\ \Eprint
  {http://arxiv.org/abs/2101.04070} {arXiv:2101.04070 [astro-ph.CO]}
  \BibitemShut {NoStop}%
\bibitem [{\citenamefont {Xu}\ \emph {et~al.}(2018)\citenamefont {Xu},
  \citenamefont {Dvorkin},\ and\ \citenamefont {Chael}}]{Xu:2018efh}%
  \BibitemOpen
  \bibfield  {author} {\bibinfo {author} {\bibfnamefont {Weishuang~Linda}\
  \bibnamefont {Xu}}, \bibinfo {author} {\bibfnamefont {Cora}\ \bibnamefont
  {Dvorkin}}, \ and\ \bibinfo {author} {\bibfnamefont {Andrew}\ \bibnamefont
  {Chael}},\ }\bibfield  {title} {\enquote {\bibinfo {title} {{Probing sub-GeV
  Dark Matter-Baryon Scattering with Cosmological Observables}},}\ }\href
  {\doibase 10.1103/PhysRevD.97.103530} {\bibfield  {journal} {\bibinfo
  {journal} {Phys. Rev. D}\ }\textbf {\bibinfo {volume} {97}},\ \bibinfo
  {pages} {103530} (\bibinfo {year} {2018})},\ \Eprint
  {http://arxiv.org/abs/1802.06788} {arXiv:1802.06788 [astro-ph.CO]}
  \BibitemShut {NoStop}%
\bibitem [{\citenamefont {Chen}\ \emph {et~al.}(2002)\citenamefont {Chen},
  \citenamefont {Hannestad},\ and\ \citenamefont {Scherrer}}]{Chen:2002yh}%
  \BibitemOpen
  \bibfield  {author} {\bibinfo {author} {\bibfnamefont {Xue-lei}\ \bibnamefont
  {Chen}}, \bibinfo {author} {\bibfnamefont {Steen}\ \bibnamefont {Hannestad}},
  \ and\ \bibinfo {author} {\bibfnamefont {Robert~J.}\ \bibnamefont
  {Scherrer}},\ }\bibfield  {title} {\enquote {\bibinfo {title} {{Cosmic
  microwave background and large scale structure limits on the interaction
  between dark matter and baryons}},}\ }\href {\doibase
  10.1103/PhysRevD.65.123515} {\bibfield  {journal} {\bibinfo  {journal} {Phys.
  Rev. D}\ }\textbf {\bibinfo {volume} {65}},\ \bibinfo {pages} {123515}
  (\bibinfo {year} {2002})},\ \Eprint {http://arxiv.org/abs/astro-ph/0202496}
  {arXiv:astro-ph/0202496} \BibitemShut {NoStop}%
\bibitem [{\citenamefont {Buen-Abad}\ \emph {et~al.}(2022)\citenamefont
  {Buen-Abad}, \citenamefont {Essig}, \citenamefont {McKeen},\ and\
  \citenamefont {Zhong}}]{Buen-Abad:2021mvc}%
  \BibitemOpen
  \bibfield  {author} {\bibinfo {author} {\bibfnamefont {Manuel~A.}\
  \bibnamefont {Buen-Abad}}, \bibinfo {author} {\bibfnamefont {Rouven}\
  \bibnamefont {Essig}}, \bibinfo {author} {\bibfnamefont {David}\ \bibnamefont
  {McKeen}}, \ and\ \bibinfo {author} {\bibfnamefont {Yi-Ming}\ \bibnamefont
  {Zhong}},\ }\bibfield  {title} {\enquote {\bibinfo {title} {{Cosmological
  constraints on dark matter interactions with ordinary matter}},}\ }\href
  {\doibase 10.1016/j.physrep.2022.02.006} {\bibfield  {journal} {\bibinfo
  {journal} {Phys. Rept.}\ }\textbf {\bibinfo {volume} {961}},\ \bibinfo
  {pages} {1--35} (\bibinfo {year} {2022})},\ \Eprint
  {http://arxiv.org/abs/2107.12377} {arXiv:2107.12377 [astro-ph.CO]}
  \BibitemShut {NoStop}%
\bibitem [{\citenamefont {McCammon}\ \emph {et~al.}(2002)\citenamefont
  {McCammon} \emph {et~al.}}]{McCammon:2002gb}%
  \BibitemOpen
  \bibfield  {author} {\bibinfo {author} {\bibfnamefont {Dan}\ \bibnamefont
  {McCammon}} \emph {et~al.},\ }\bibfield  {title} {\enquote {\bibinfo {title}
  {{A High spectral resolution observation of the soft x-ray diffuse background
  with thermal detectors}},}\ }\href {\doibase 10.1086/341727} {\bibfield
  {journal} {\bibinfo  {journal} {Astrophys. J.}\ }\textbf {\bibinfo {volume}
  {576}},\ \bibinfo {pages} {188--203} (\bibinfo {year} {2002})},\ \Eprint
  {http://arxiv.org/abs/astro-ph/0205012} {arXiv:astro-ph/0205012} \BibitemShut
  {NoStop}%
\bibitem [{\citenamefont {Erickcek}\ \emph {et~al.}(2007)\citenamefont
  {Erickcek}, \citenamefont {Steinhardt}, \citenamefont {McCammon},\ and\
  \citenamefont {McGuire}}]{Erickcek:2007jv}%
  \BibitemOpen
  \bibfield  {author} {\bibinfo {author} {\bibfnamefont {Adrienne~L.}\
  \bibnamefont {Erickcek}}, \bibinfo {author} {\bibfnamefont {Paul~J.}\
  \bibnamefont {Steinhardt}}, \bibinfo {author} {\bibfnamefont {Dan}\
  \bibnamefont {McCammon}}, \ and\ \bibinfo {author} {\bibfnamefont
  {Patrick~C.}\ \bibnamefont {McGuire}},\ }\bibfield  {title} {\enquote
  {\bibinfo {title} {{Constraints on the Interactions between Dark Matter and
  Baryons from the X-ray Quantum Calorimetry Experiment}},}\ }\href {\doibase
  10.1103/PhysRevD.76.042007} {\bibfield  {journal} {\bibinfo  {journal} {Phys.
  Rev. D}\ }\textbf {\bibinfo {volume} {76}},\ \bibinfo {pages} {042007}
  (\bibinfo {year} {2007})},\ \Eprint {http://arxiv.org/abs/0704.0794}
  {arXiv:0704.0794 [astro-ph]} \BibitemShut {NoStop}%
\bibitem [{\citenamefont {Mahdawi}\ and\ \citenamefont
  {Farrar}(2018)}]{Mahdawi:2018euy}%
  \BibitemOpen
  \bibfield  {author} {\bibinfo {author} {\bibfnamefont {M.~Shafi}\
  \bibnamefont {Mahdawi}}\ and\ \bibinfo {author} {\bibfnamefont {Glennys~R.}\
  \bibnamefont {Farrar}},\ }\bibfield  {title} {\enquote {\bibinfo {title}
  {{Constraints on Dark Matter with a moderately large and velocity-dependent
  DM-nucleon cross-section}},}\ }\href {\doibase 10.1088/1475-7516/2018/10/007}
  {\bibfield  {journal} {\bibinfo  {journal} {JCAP}\ }\textbf {\bibinfo
  {volume} {10}},\ \bibinfo {pages} {007} (\bibinfo {year} {2018})},\ \Eprint
  {http://arxiv.org/abs/1804.03073} {arXiv:1804.03073 [hep-ph]} \BibitemShut
  {NoStop}%
\bibitem [{\citenamefont {Genzel}\ \emph {et~al.}(2010)\citenamefont {Genzel},
  \citenamefont {Eisenhauer},\ and\ \citenamefont {Gillessen}}]{Genzel:2010zy}%
  \BibitemOpen
  \bibfield  {author} {\bibinfo {author} {\bibfnamefont {Reinhard}\
  \bibnamefont {Genzel}}, \bibinfo {author} {\bibfnamefont {Frank}\
  \bibnamefont {Eisenhauer}}, \ and\ \bibinfo {author} {\bibfnamefont {Stefan}\
  \bibnamefont {Gillessen}},\ }\bibfield  {title} {\enquote {\bibinfo {title}
  {{The Galactic Center Massive Black Hole and Nuclear Star Cluster}},}\ }\href
  {\doibase 10.1103/RevModPhys.82.3121} {\bibfield  {journal} {\bibinfo
  {journal} {Rev. Mod. Phys.}\ }\textbf {\bibinfo {volume} {82}},\ \bibinfo
  {pages} {3121--3195} (\bibinfo {year} {2010})},\ \Eprint
  {http://arxiv.org/abs/1006.0064} {arXiv:1006.0064 [astro-ph.GA]} \BibitemShut
  {NoStop}%
\bibitem [{\citenamefont {Schoedel}\ \emph {et~al.}(2008)\citenamefont
  {Schoedel}, \citenamefont {Merritt},\ and\ \citenamefont
  {Eckart}}]{Schoedel:2008ny}%
  \BibitemOpen
  \bibfield  {author} {\bibinfo {author} {\bibfnamefont {R.}~\bibnamefont
  {Schoedel}}, \bibinfo {author} {\bibfnamefont {D.}~\bibnamefont {Merritt}}, \
  and\ \bibinfo {author} {\bibfnamefont {A.}~\bibnamefont {Eckart}},\
  }\bibfield  {title} {\enquote {\bibinfo {title} {{The nuclear star cluster of
  the Milky Way}},}\ }\href {\doibase 10.1088/1742-6596/131/1/012044}
  {\bibfield  {journal} {\bibinfo  {journal} {J. Phys. Conf. Ser.}\ }\textbf
  {\bibinfo {volume} {131}},\ \bibinfo {pages} {012044} (\bibinfo {year}
  {2008})},\ \Eprint {http://arxiv.org/abs/0810.0204} {arXiv:0810.0204
  [astro-ph]} \BibitemShut {NoStop}%
\bibitem [{\citenamefont {Baganoff}\ \emph {et~al.}(2003)\citenamefont
  {Baganoff} \emph {et~al.}}]{Baganoff:2001ju}%
  \BibitemOpen
  \bibfield  {author} {\bibinfo {author} {\bibfnamefont {Frederick~K.}\
  \bibnamefont {Baganoff}} \emph {et~al.},\ }\bibfield  {title} {\enquote
  {\bibinfo {title} {{Chandra x-ray spectroscopic imaging of Sgr A* and the
  central parsec of the Galaxy}},}\ }\href {\doibase 10.1086/375145} {\bibfield
   {journal} {\bibinfo  {journal} {Astrophys. J.}\ }\textbf {\bibinfo {volume}
  {591}},\ \bibinfo {pages} {891--915} (\bibinfo {year} {2003})},\ \Eprint
  {http://arxiv.org/abs/astro-ph/0102151} {arXiv:astro-ph/0102151} \BibitemShut
  {NoStop}%
\end{thebibliography}%

\end{document}